\documentclass[reprint,aps,prl,superscriptaddress]{revtex4-2}

\usepackage{amsmath, amssymb, times, mathrsfs, hyperref, array, bbm}
\usepackage{graphicx}
\usepackage[usenames,dvipsnames]{xcolor}
\usepackage{braket}
\usepackage{tcolorbox}
\usepackage{listings}
\hypersetup{colorlinks=false}
\usepackage{ulem}
\usepackage{balance}

\begin{document}
\title{Flux-tunable global and local superconductivity in a topological insulator nano-SQUID}

\author{Ella Nikodem}
\affiliation{Physics Institute II, University of Cologne, Z\"ulpicher Straße 77, 50937 K\"oln, Germany}

\author{Jakob Schluck}
\affiliation{Physics Institute II, University of Cologne, Z\"ulpicher Straße 77, 50937 K\"oln, Germany}

\author{Micha{\l} Papaj}
\affiliation{Department of Physics, University of Houston, Houston, TX 77204, USA}

\author{Max Geier}
\affiliation{Department of Physics, Massachusetts Institute of Technology, Cambridge MA 02139, USA}

\author{Mahasweta Bagchi}
\affiliation{Physics Institute II, University of Cologne, Z\"ulpicher Straße 77, 50937 K\"oln, Germany}

\author{Liang Fu}
\affiliation{Department of Physics, Massachusetts Institute of Technology, Cambridge MA 02139, USA}

\author{Henry~F.~Legg}
\affiliation{SUPA, School of Physics and Astronomy, University of St Andrews,
North Haugh, St Andrews, KY16 9SS, United Kingdom}

\author{Yoichi Ando}
\affiliation{Physics Institute II, University of Cologne, Z\"ulpicher Straße 77, 50937 K\"oln, Germany}
\maketitle

\textbf{Topological systems are defined by global properties that enforce the existence of local boundary modes. Three-dimensional topological insulators (TIs) were among the earliest proposed systems for hosting topological superconductivity, but experimental focus subsequently shifted to other platforms. Here, we revisit bulk-insulating TIs using a columnar nano-superconducting quantum interference device (nano-SQUID) architecture. This geometry optimises the proximity effect on the TI surface and enables simultaneous probing of global superconducting properties --- via the critical current through the nano-SQUID --- alongside the local states at the ends of the nano-SQUID via tunnel junctions. We observe several global superconducting features that appear to show a flux-driven global phase transition consistent with entering the topological regime, including periodic critical current oscillations and a sign reversal in the superconducting diode effect. Simultaneously, tunnelling spectroscopy reveals spectral jumps in local and nonlocal conductance that align with these global features. However, zero-bias peaks (ZBPs) in local conductance are present both within the predicted topological range of magnetic fields and in theoretically trivial regimes, including at zero magnetic field. {Ultimately, the lack of correlation between local ZBP signatures and global signatures emphasises that conclusively identifying Majorana bound states (MBSs) will necessitate a combined approach, integrating the establishment of global topological properties with the use of local and other, more advanced, probes.}}


The search for topological superconductivity has long been motivated by the prediction that such a phase hosts non-Abelian MBSs, the potential building blocks for fault-tolerant quantum computing~\cite{Nayak2008}. Three-dimensional topological insulators (TIs) were among the first material classes identified as candidates for this pursuit due to their {inherent} spin-momentum locked surface states~\cite{Fu2008}. Theoretically, superconducting TIs should host topological superconductivity if the chemical potential is tuned within the bulk band gap ($\sim$300~meV in the BiSbTeSe$_2$ devices considered here~\cite{Arakane2012}). However, TIs are not usually intrinsically superconducting, necessitating the use of the superconducting proximity effect, which complicated efforts to find MBSs~\cite{Breunig2021}. Furthermore, as in other platforms, distinguishing MBSs in the potential topological phase from trivial bound states has proved persistently difficult~\cite{Yazdani2023}.

Following the initial interest in TIs, experimental focus largely shifted to semiconductor nanowires with strong spin-orbit coupling, such as InAs or InSb~\cite{Mourik2012,Albrecht2016,Deng2016,Vaitiekenas2020,Aghaee2025}. In these systems, the topological transition is predicted to be achieved by the combination of Zeeman energy from an applied magnetic field and electrostatic gating, which allows access to the resulting helical regime in the normal state. Despite extensive efforts, conclusive evidence of topological superconductivity in such systems remains elusive, in part because disorder effects frequently mimic topological signatures~\cite{Valentini2021,Hess2023}. Even in the ideal case, the topological regime in such a setup is limited to a few 100~$\mu$eV range of chemical potential, making the phase fragile to local potential fluctuations. The resulting ambiguity has necessitated complex tune-up protocols merely to identify a purported topological phase~\cite{Microsoft2023,Legg2026}. To address this, recent strategies have explored artificial Kitaev chains in quantum dot arrays to mitigate the potential disorder landscape via gate control~\cite{Dvir2023,tenHaaf2024,tenHaaf2025,vanLoo2026}, but such efforts are still in development and even in ideal theoretical scenarios, they realise only small topological regions of phase space~\cite{Luethi2024,Luethi2025}.

\begin{figure*}[htbp]
\includegraphics{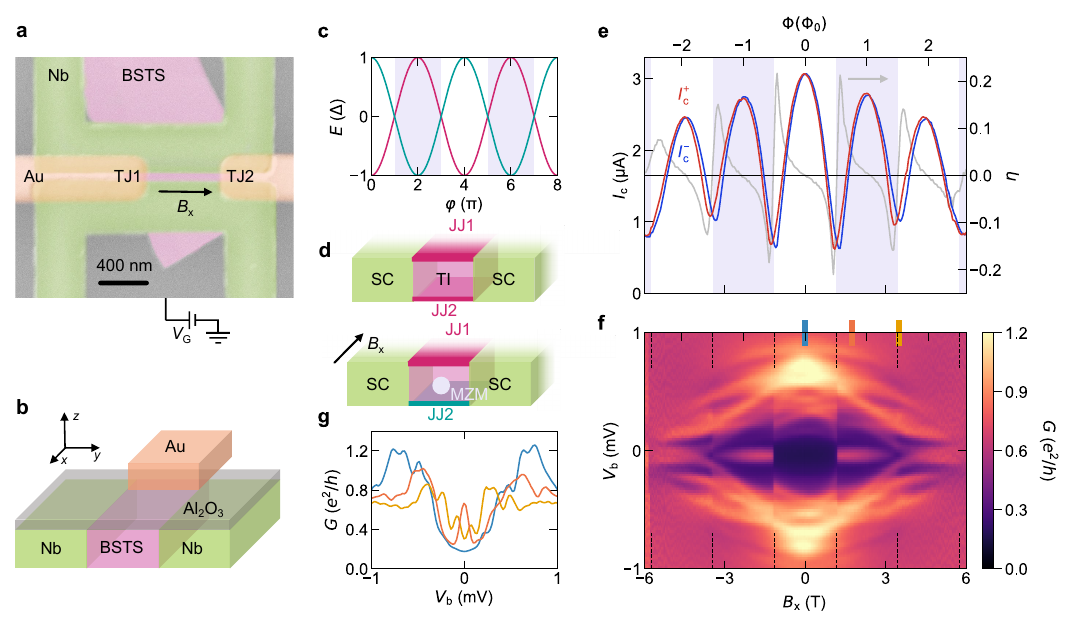}
  \caption{\textbf{Flux-tunable transport and local spectroscopy in a nano-SQUID.} \textbf{a}, Scanning electron microscope (SEM) image of the device. An exfoliated topological-insulator (TI) flake is patterned into a nanowire (pink) by dry etching. The etched regions are filled with Nb (green), forming a sandwich junction. Residual TI fragments do not participate in transport. Gold electrodes contact the TI nanowire via tunnel junctions TJ1 and TJ2, enabling local probing of the density of states. \textbf{b}, Schematic of the baseline device, featuring the TI nanowire core, the superconducting shell, tunnel dielectric, and a single functional TJ. \textbf{c}, Energy-phase dispersion of the perpendicular {($k_x=0$)} mode in a TI line junction. \textbf{d}, Schematic of the effective nano-SQUID formed by the two Josephson junctions, JJ1 and JJ2, located on the top and bottom surfaces of a bulk-insulating TI nanowire. An in-plane magnetic field induces different local phase differences across the two junctions, allowing the top and bottom surfaces to enter different topological regimes. Majorana zero modes (MZMs) can emerge at the resulting phase boundaries. \textbf{e}, $I_c^+$ (red) and $I_c^-$ (blue) of device A measured as a function of the axial magnetic field $B_x$. To avoid hysteresis effects, the current was always ramped up from zero. The SDE efficiency $\eta$ (grey) shows sign reversals at half-integer flux quanta. Expected topological regions are highlighted in coloured columns. \textbf{f,} $G$ as a function of $V_\text{b}$ and $B_x$, at the single TJ of device A for a backgate voltage of $V_\text{G}=10~$V. Markers along the top axis indicate the magnetic field values for which the line cuts presented in panel \textbf{g} were extracted. \textbf{g}, Line cuts of $G$ highlighting spectral gap evolution at different fluxes.}\label{Fig1}
\end{figure*}

Here, we revisit the foundational promise of the 3D TI using the recently developed columnar nano-superconducting quantum interference device (nano-SQUID) geometry (see Fig.~\ref{Fig1}a,b)~\cite{Nikodem2025}. This platform, in which a thin nanowire of a TI is sandwiched between two superconductors to form a lateral Josephson junction, combines the intrinsic robustness of the TI bulk gap with superconducting phase control.  The topological transition in this architecture is driven by a superconducting phase difference across the nanowire rather than electrostatic gating or the Zeeman effect. Importantly, the nano-SQUID effectively forms two superconducting-normal-superconducting (SNS) line junctions on the top and bottom surfaces, each undergoing topological phase transitions controlled by their superconducting phase difference (Fig.~\ref{Fig1}c)\cite{Fu2008,Schluck2024}. This phase difference is controlled externally by an axial magnetic flux, $\Phi$, through the nanowire cross-section (Fig.~\ref{Fig1}d).  In this setup, the topological phase is centered around an odd number of superconducting flux quanta ($\Phi_0=h/2e$) threading the system~\cite{Nikodem2025}. In this regime, it becomes energetically favorable for the superconducting ground state phase to undergo a phase jump of $\pi$, whereupon the system enters the topological regime. Importantly, since this defines the lowest energy configuration of the superconducting phase difference, the topological superconducting phase is self-sustaining within a broad range of magnetic flux around this value, unlike the case for proposals based on a single junction~\cite{Pientka2017,Hell2017,Dartiailh2021}. Perhaps most significantly compared to many other platforms, the topological insulator nano-SQUID architecture permits the simultaneous probing of global and local features of the superconducting state through, respectively, the Josephson junctions critical current and the use of tunnel probes at the ends of the nanowire in the nano-SQUID.

\begin{figure}[t]
\includegraphics[width=\columnwidth]{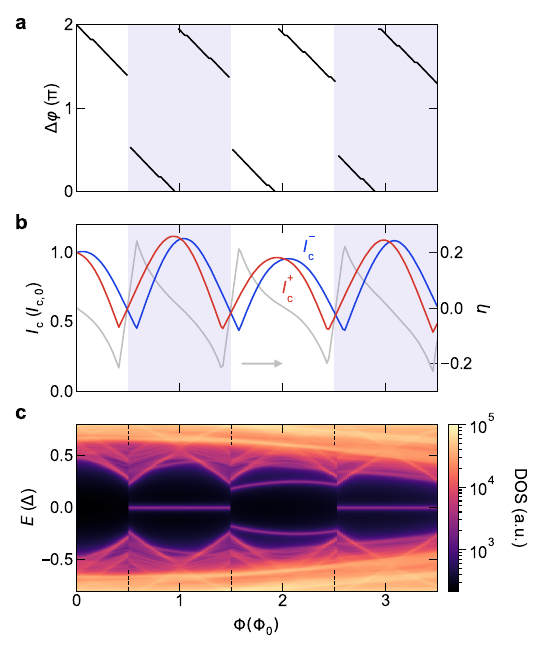}
  \caption{\textbf{Theory of topological superconductivity in a TI nano-SQUID.}  \textbf{a,} Calculated ground-state phase difference, explicitly mapping the $0-\pi$ transition to topological boundaries. \textbf{b,} Calculated $I_c$ and $\eta$ vs. flux $\Phi$. \textbf{c,} Simulated energy spectrum as a function of $\Phi$, showing the spectral jumps at the values of the topological phase transition and the emergence of zero-energy modes in regions with an odd number of flux quanta.}\label{Fig2}
\end{figure}

In Fig.~\ref{Fig1} we present the data from a first-generation device (device A) featuring a single functional tunnel contact. To probe the global properties of this TI nano-SQUID, we measure the critical currents across the junction for positive bias, $I_c^+$, and negative bias, $I_c^-$, as a function of the parallel magnetic field $B_x$ (Fig.~\ref{Fig1}e). $I_c$ exhibits strong oscillations with a periodicity corresponding to one flux quantum $\Phi_0 = h/2e$ threading the effective cross-sectional area of the nanowire (Fig.~\ref{Fig1}b and supplemental note S1.A). This $h/2e$ periodicity and the almost complete revival of $I_c$ for an integer flux quantum demonstrate that the surface state is dominant such that there is a negligible bulk contribution to the supercurrent \cite{Nikodem2025}. Simultaneously with these $I_c$ oscillations, we observe a flux-tunable superconducting diode effect (SDE) where the $I_c$ depends on the direction of current flow ($I_c^+ \neq I_c^-$)\cite{Ando2020, Baumgartner2022, Nikodem2025diode}. As shown in Fig.~\ref{Fig1}e, this SDE reaches an efficiency of $\eta \approx 20\%$, where $\eta = (I_c^+ - I_c^-)/ (I_c^+ + I_c^-)$, similar to that found previously\cite{Nikodem2025diode}. An important property of this SDE is that its polarity reverses at half-integer flux quanta, $\Phi \approx \Phi_0(n+1/2)$, with $n$-integer \cite{Nikodem2025diode}. The uniform phase oscillations in $I_c$, presence of the highly efficient SDE, and the reversal of the SDE exactly at half-integer flux quanta, all fit the expected theoretical global signatures of a ground-state phase experiencing a phase jump across the junction, which theoretically corresponds to the transition from trivial to topological superconductivity (see Fig.~\ref{Fig2}a-b) \cite{Nikodem2025, Nikodem2025diode}.

Local differential conductance measurements (Fig.~\ref{Fig1}b, ~\ref{Fig1}f) allow us to relate these global features to the microscopic density of states (see Fig.~\ref{Fig1}g). As the magnetic flux is tuned, the superconducting gap edge and sub-gap states --- Andreev Bound States (ABSs) --- exhibit strong dispersion. In this first generation device with a single functioning tunnel probe, at the flux corresponding to the diode polarity reversal and theoretical phase transition to topological superconductivity, we observe a jump in the spectrum and the appearance of a zero-bias peak (ZBP) within the predicted topological window. As $B_x$ is further increased the ZBP persists for a large range of $B_x$. The ZBP finally disappears and a state split from zero-bias appears in the regime where two flux quanta thread the nanowire, which theory predicts should be trivial. The phenomenology remains qualitatively identical for different chemical potentials, as tuned by the global back gate voltage (see Supplementary Note S1.D).

\begin{figure}[htbp]
  \centering
\includegraphics[width=\columnwidth]{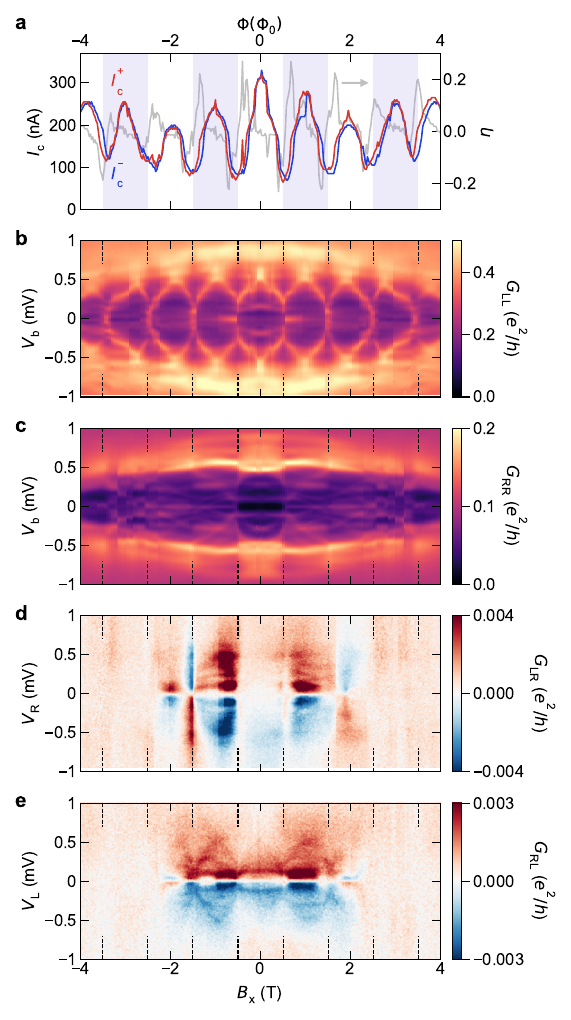}
  \caption{\textbf{Simultaneous global and dual-local spectroscopy.} \textbf{a,} Global $I_c$ and $\eta$ in device B. Note that the cross-section is larger than for device A and multiple oscillations of $I_c$ are observed. \textbf{b, c,} Differential conductance measured simultaneously at TJ1 ($G_{LL}$, b) and TJ2 ($G_{RR}$, c). Abrupt spectral jumps occur at fields commensurate with $I_c$ anomalies. However, there are zero-bias peaks visible in both theoretically topological regimes and trivial regimes.  \textbf{d, e,} Nonlocal conductance $G_{LR}$ and $G_{RL}$ showing that some features correspond to the boundaries of the expected topological regimes.}
\end{figure}

\begin{figure*}[t]
  \centering
\includegraphics[width=\textwidth]{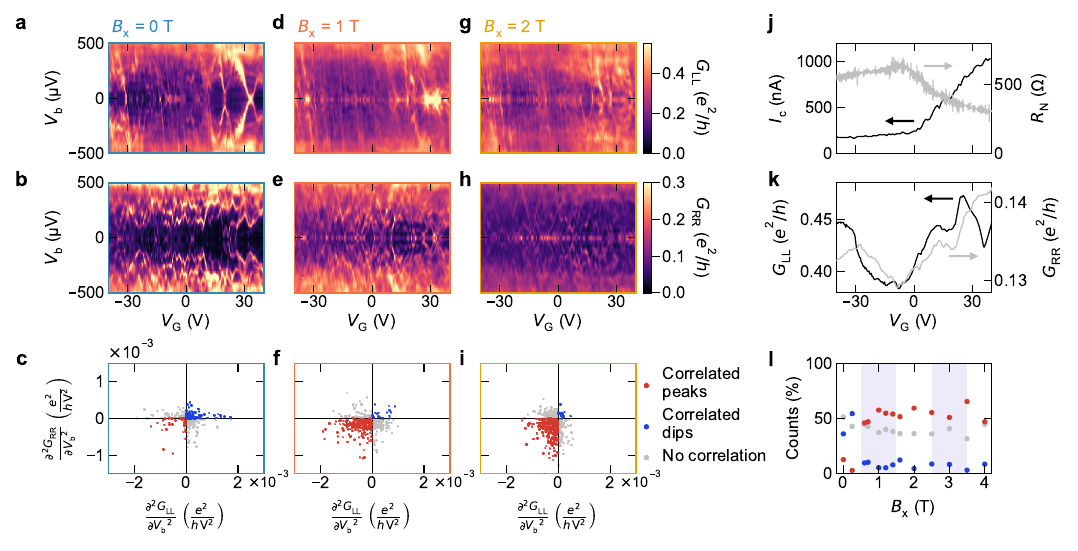}
  \caption{\textbf{Gate-voltage dependence and end-to-end correlations.} \textbf{a, b, d, e, g, h} $G_{LL}$ and $G_{RR}$ in device B as a function of $V_\text{G}$ for trivial and topological flux regimes. \textbf{c, f, i} End-to-end correlation analysis using the product of the the third derivative at zero bias. \textbf{j} $V_\text{G}$-dependence of $I_c$ (black) and the normal state resistance $R_N$ (grey). \textbf{k} $V_\text{G}$-dependence of $G_{LL}$ and $G_{RR}$. \textbf{l} Relative fraction of gate voltages exhibiting correlated zero-bias conductance peaks (red), correlated zero-bias conductance dips (blue), or no correlation (grey) as a function of $B_x$.}\label{Fig4}
\end{figure*}
The results in this device are all broadly consistent with the theoretical expectation that there is a topological superconducting phase centered around an odd number of superconducting flux quanta. In Fig.~2 we present the ideal theoretical situation, which is calculated using a three-dimensional tight binding model (see Methods). As the magnetic flux is increased, the phase difference across the junction undergoes a $\pi$ phase jump at half-integer values of flux quanta (Fig.~2a). Correspondingly, theory predicts an abrupt change in the sign of the SDE and a recovery of the $I_c$ for integer values of flux quanta (see Fig.~2b), as observed in our devices. Furthermore, turning to the energy of individual states (Fig.~2c), theory predicts that finite energy ABSs oscillate in phase with the phase jumps across the junction, such that the spectrum is approximately restored for integer values of flux quanta. Finally, due to the topological nature of the system when odd-integer flux quanta thread it, zero-energy MBSs occur at the end of the nanowire resulting in ZBPs in the tunnelling spectra on either side of the nanowire.

Encouraged that both the local and global features predicted by theory are observed in this first generation device, we improved the fabrication and were able to produce a series of devices with two functioning tunnel contacts. The implementation of a dual-probe architecture facilitates a direct comparison between global properties and microscopic boundary states at both ends of the nanowire. As in the first generation device, the global signatures of the topological phase transition --- specifically $I_c$ oscillations and the sign reversal of the superconducting diode effect --- are also observed in all examined devices and are largely robust against gate voltage (see Supplemental Note S1.F). For instance the device in Fig.~3a (device B), which has an approximately 70\% larger cross-section than the device in Fig.~1, exhibits multiple $I_c$ oscillations and corresponding changes in the SDE. In Fig.~3b and c we show the differential conductance simultaneously detected at both ends of the TINW as $B_x$ is tuned, with the left-end local conductance $G_\text{LL}$ measured via tunnel junction 1 (TJ1) and the right-end conductance $G_\text{RR}$ measured via tunnel junction 2 (TJ2).  Both ends show a rich spectrum of ABSs evolving with the field, showing characteristic discontinuities correlating with the minima in supercurrent and SDE efficiency, consistent with the findings for device A and an additional device C (see Supplementary Note S1.E). This consistency across a number of devices strongly supports a flux-driven global phase transition governed by the phase difference across the nano-SQUID. 

At high magnetic fields $B_x$, the behaviour is complicated by the suppressed SC gap of the parent Nb electrodes, which consequently reduces the induced gap of the proximitized TINW. Moreover, at high magnetic field, irregular jumps appear in the conductance spectra at magnetic fields that do not coincide with flux values of $\Phi = \Phi_0 (n + \frac{1}{2})$, where $n$ is an integer. Their strong dependence on the magnetic-field sweep direction points to a vortex dynamics and flux trapping origin within the Nb electrodes. Because the nominal in-plane field $B_x$ inevitably acquires a finite out-of-plane component that grows linearly with its magnitude, even a slight misalignment of $B_x$ relative to the device plane introduces an increasing effective $B_z$, further facilitating vortex entry into the Nb leads.

Additionally, with two tunnel probes, we can measure the nonlocal conductance through the length of the nanowire, which in principle adds another global probe of the superconductivity\cite{Menard2020,Pan2021,Hess2021,Poschl2022,Banerjee2023PRL,Hess2023}. We define the differential conductance matrix elements as $G_{ij} = dI_i/dV_j$, where $I_i$ is the current measured at tunnel junction $i$ in response to a voltage excitation $V_j$ applied to tunnel junction $j$, where $i,j\in \{\text{L,R}\}$. Interestingly, the nonlocal conductance reveals features that align with the $I_c$ oscillations, such as jumps at the boundaries where the phase transition is supposed to occur (see Fig.~3d-e). The coincidence of nonlocal conductance features and those from $I_c$ is broadly found in all devices. However, whilst features in the nonlocal conductance do often coincide with the $I_c$ features, there are also nonlocal conductance jumps at $B_x$ values that do not correspond to the $I_c$ features, especially at large values of magnetic flux. This likely indicates the complex interplay of effects local to the tunnel contacts, such as local phase instabilities, and global properties of the full device that impact the nonlocal conductance.

Despite the clear theoretical expectation and experimental global signatures of a flux-induced phase transition to topological superconductivity, local tunnel spectroscopy is substantially more ambiguous. We do find that the subgap states often exhibit jumps close to the expected phase transition point, similar to those exhibited in the first generation devices (see Fig.~3b-c). However, focussing on the ZBPs that should correspond to MBSs in these devices, we observe these even in theoretically trivial windows and notably even at zero $B_x$. Although the zero field ZBPs largely have a strong dependence on $V_\text{G}$ (Fig.~4a-b) we do not see a significant difference between the $V_\text{G}$ dependence of ZBPs in the expected topological or trivial regimes (Fig.~4d-e \& g-h) even over a region of $V_\text{G}$ that includes the Dirac point (Fig.~4j-k).

To further clarify whether the observed ZBPs might be topological, we analysed the correlation of local states at opposite ends of the nanowire. If these ZBPs originated from Majorana bound states, we would expect a positive correlation i.e. a ZBP on either end of the nanowire in the nano-SQUID. Given the induced superconducting gap and the typical Fermi velocity~\cite{Arakane2012} of BiSbTeSe$_2$ surface states, we estimate the experimental localisation length of the MBSs to be a few 100~nm, which is shorter than the  $1.3-2.5\mu$m length of our devices (see Supplementary Note S1.A). Hence, the boundary modes should be localised at the junctions without significant overlap in our devices. Surprisingly our measurements consistently show that there is no clear distinction between expected topological regions and trivial regions, with both showing a large number of ZBPs at both ends. To quantify this, we evaluated the end-to-end correlation using the relative signs of the third derivative of the current at zero bias: $\mathcal{C}_{LR} \propto \mathrm{sgn}(\partial^3 I_L/\partial V_L^3) \times \mathrm{sgn}(\partial^3 I_R/\partial V_R^3)$.  We find that both regimes contain ZBPs on either end (see Fig.~4c,f,i,l) the majority of the time. The fact this occurs even in theoretically trivial regions suggests that at least some of these ZBPs are driven by disorder effects, rather than a potential topological superconducting phase.

The critical current oscillations, diode effect reversals, and local spectral jumps all match the theoretical expectation for the realisation of a  robust topological superconducting phase in our TI nano-SQUID architecture. In stark contrast, end-to-end correlations of zero-bias peaks yield similar results within both the predicted topological and trivial regimes within the higher lobes, showing no clear relation to the corresponding signatures of the bulk phase. This disparity highlights that an apparently clear bulk topological transition does not guarantee the isolation of topological boundary modes. Consequently, even when global signatures strongly suggest a robust bulk topological state is successfully realised, as is the case in the topological nano-SQUID investigated here, there remains an experimental bottleneck to isolate true Majorana modes from trivial states that obscure them. Our findings demonstrate that new platforms, where global properties can be assessed simultaneously with local features, are an essential step forward in the search for Majorana bound states, but that a conclusive demonstration will also require the development of advanced probes of Majorana physics --- with electron teleportation \cite{Fu2010} serving as one potential candidate --- within regimes that exhibit global topological signatures.

{\bf  Methods:}

{\bf  Materials and device fabrication:} Single crystals of BiSbTeSe$_2$ were synthesized from 99.9999\% pure Bi, Sb, Te, and Se using a modified Bridgman growth method, following the procedure described in Ref. \cite{Ren2011}. Thin BiSbTeSe$_2$ flakes (thickness $d\sim20$ nm) were obtained by mechanical exfoliation and transferred onto doped Si substrates covered with a 290 nm SiO$_2$ layer, which served as the back-gate dielectric. Suitable flakes for device fabrication were selected using optical microscopy. Device patterning was performed using a Raith PIONEER Two system for electron-beam lithography. In a single electron-beam lithography step both the TINWs and the junction electrodes were defined. The BiSbTeSe$_2$ flakes were first etched into  rectangular nanowires with a width $W<100$ nm and a length $L$ of a couple of micrometers by Ar plasma dry etching. Subsequently, a 45 nm layer of Nb ($T_\text{c}\sim7$ K and $H_\text{c2,in-plane}\gtrsim 6$ T) was sputter-deposited to form the superconducting junctions. To electrically isolate the superconducting electrodes from the tunnel probes fabricated in later processing steps, the Nb was coated with a few nanometers of Al$_2$O$_3$ layer grown by atomic layer deposition (ALD) using an Ultratec Savannah S200 system. Following the lift-off process, a second ALD process was used to deposit an additional ultrathin Al$_2$O$_3$ layer ($\leq1$ nm) over the entire device. This layer serves as tunnel barrier. The tunnel electrodes were defined in another electron-beam lithography step. A 5 nm Pt wetting layer and a 45 nm Au layer were sputter-deposited to form the tunnel electrodes. 

The resulting tunnel contacts typically exhibit resistances exceeding 100 k$\Omega$, ensuring that electron tunneling dominates the transport. Although Josephson junction fabrication is highly reproducible, the yield of functional tunnel junctions fabricated on top remains below 50\%, making devices with two decent tunnel probes to be comparatively rare.  This limitation primarily arises from the extremely small thickness required for the dielectric tunnel barrier and the high sensitivity of the device performance to thickness variations in either directions. 

After completing the transport measurements, the precise device dimensions were determined by SEM. 

{\bf  Measurements:} Electrical characterization of the Josephson junctions was performed in a dry dilution refrigerator (Oxford Instruments TRITON 300) operating at a base temperature of approximately 30 mK. To suppress electrical noise, all measurement lines were equipped with RC filters and copper-powder filters.

The differential resistance, d$V$/d$I$, was generally measured in a quasi-four-terminal configuration. In cases where broken leads prevented such setup, a two-terminal configuration was employed and the known filter resistance was subtracted from the recorded data. A standard low-frequency lock-in technique with DC bias was applied to measure the differential resistance. 

Tunnel spectroscopy measurements were performed in a two-terminal setup between each individual tunnel junction and a superconducting electrode. The differential conductance, $G$, was obtained using a standard low-frequency lock-in technique in combination with a Basel Instrument current amplifier. A small DC offset introduced by the preamplifier was numerically corrected for during data analysis. 

For nonlocal conductance measurements, independent AC modulations at frequencies $f_\text{L}$ and $f_\text{R}$ were applied to the left and right tunnel junctions, respectively. Lock-in detection at $f_\text{L}$ was used to extract the conductance associated with the left-side excitation, while detection at $f_\text{R}$ provided the corresponding response to the right-side modulation.  In this way, both local and nonlocal conductance components could be determined. Details of the experimental setup and the analysis of the nonlocal conductance data can be found in Ref. \cite{Feng2024} and the Supplementary Information. 

Magnetic-field-dependent measurements were carried out using a 6/1/1 T superconducting vector magnet, allowing precise alignment of the magnetic field relative to the nanowire axis. A detailed description of the alignment procedure can be found in Ref. \cite{Nikodem2025}.

{\bf  Device simulations:} The device simulations using Kwant package \cite{Groth2014} are based on finite-difference lattice discretization of continuum models with three distinct regions: two superconducting electrodes described by a simple parabolic band with a superconducting order parameter, and a central region described by a three-dimensional massive Dirac fermion Hamiltonian. We note that in the simulations and the model description below, $x$ is the direction across the junction, while $y$ is the direction along the nanowire axis. Expressed in momentum space, the Hamiltonian in the central topological insulator nanowire is:

\begin{equation}
\begin{split}
    H_\mathrm{TI}(\mathbf{k}) &= \left(D_2(k_x^2 + k_y^2) + D_1 k_z^2 - \mu_\mathrm{TI} \right) s_0 \sigma_0 \\ &+ \left(M - B_2 (k_x^2 + k_y^2) - B_1 k_z^2 \right) s_0\sigma_z \\ &+ A_2 (k_x s_x \sigma_x + k_y s_y \sigma_x) + A_1 k_z s_z \sigma_x
\end{split}
\end{equation}
where $s_i$ and $\sigma_i$ are Pauli matrices in the spin and orbital subspaces, and $\mu_\mathrm{TI}$ is the chemical potential within the topological insulator nanowire. The Hamiltonian of the superconducting electrodes is then:
\begin{equation}
\begin{split}
    H_\mathrm{SC}(\mathbf{k}) =& t(k_x^2 + k_y^2 + k_z^2 - \mu_\mathrm{SC}) \tau_z s_0 \sigma_0 \\ &- \mathrm{Re}(\Delta) \tau_y s_y \sigma_0 - \mathrm{Im} (\Delta) \tau_x s_y \sigma_0
\end{split}
\end{equation}
where $\tau_i$ are the Pauli matrices in the particle-hole space, $\Delta = \Delta_0 \exp(i \phi(x))$ is the superconducting order parameter characterized by its magnitude $\Delta_0$ and phase $\phi(x)$, and $\mu_\mathrm{SC}$ is the chemical potential within the superconducting electrode. The superconducting phase is defined for the two superconducting electrodes as follows:
\begin{equation}
    \phi(x) = 
    \begin{cases} 
        -\frac{\phi}{2} & \text{for left SC electrode} \\
        +\frac{\phi}{2} & \text{for right SC electrode}
    \end{cases}
\end{equation}
Therefore, $\phi$ gives the superconducting phase difference across the Josephson junction.

To incorporate the magnetic field in the system, the Peierls' substitution was used to add the complex phase to all the hoppings $t_{ij}$ from the site at $\mathbf{r}_j$ to the site at $\mathbf{r}_i$ through:
\begin{equation}
    t_{ij} \rightarrow t_{ij} \exp\left(-i \frac{\pi}{\Phi_0} \int_{\mathbf{r}_j}^{\mathbf{r}_i}d\mathbf{r}\cdot\mathbf{A}(\mathbf{r})\right)
\end{equation}
where $\mathbf{A}(\mathbf{r})=B_y z \hat{x}$ is the vector potential in the chosen gauge that results in a magnetic field $\mathbf{B} = B_y \hat{y}$ and $\Phi_0 = h/(2e)$ is the superconducting flux quantum. Because the superconducting electrodes in the device are thin compared to the London penetration depth, it was assumed that the magnetic field fully penetrates the whole device (including the superconducting regions) and its magnitude is not affected by the Meissner effect.

To observe superconducting diode effect, it is also necessary to break inversion symmetry, which in the experiment is achieved through gating. The applied $V_\text{GT}$ and resulting electric field is included in the simulation via position-dependent electrostatic potential:
\begin{equation}
    V(z) = V_\text{GT} \frac{z}{H}
\end{equation}
where $V_\text{GT}$ is the applied potential difference between the top and bottom surfaces of the wire, and $H$ is the height of the nanowire. This electrostatic potential is included only within the topological insulator nanowire region.

Both Hamiltonians are discretized on a square lattice with lattice spacing $a_\mathrm{lat} = 1\,$nm in two configurations. The first assumes translational invariance along the nanowire axis, which means that momentum $k_y$ remains a good quantum number. In that case, the full spectrum of the system $E_i(k_y, \phi)$ was obtained and used to calculate the total energy- and current-phase relationship of the Josephson junction:
\begin{equation}
\begin{split}
    E_\mathrm{tot}(B_y, \phi) &= \sum_{E_i<0, k_y} E_i(B_y, k_y, \phi), \\ \quad I_\mathrm{tot}(B_y, \phi) &= \frac{d E_\mathrm{tot}(B_y, \phi)}{d\phi}.
\end{split}
\end{equation}
For each value of magnetic field (and thus the magnetic flux threaded through the device), the critical currents were then determined in both directions:
\begin{equation}
\begin{split}
    I_{c+}(B_y) &= \max_{\phi \in [0, 2\pi]} I_\mathrm{tot}(B_y, \phi), \\ \quad I_{c-}(B_y) &= \min_{\phi \in [0, 2\pi]} I_\mathrm{tot}(B_y, \phi).
\end{split}
\end{equation}
These critical supercurrents were then used to determine the superconducting diode efficiency defined as:
\begin{equation}
    \eta = \frac{I_{c+} - |I_{c-}|}{I_{c+} + |I_{c-}|}.
\end{equation}
These superconducting diode simulation results are then presented in Fig.~\ref{Fig2}(b). This translationally-invariant system configuration was also used to determine superconducting phase differences $\phi_\mathrm{min}(B_y)$ for which no net supercurrent is carried across the junction, and which correspond to the global minimum of total energy, at each magnetic field. These minimum phases exhibited discontinuities at half-odd-integer superconducting flux quanta threaded through the device. 

The second considered configuration was a full three-dimensional lattice with a nanowire of finite length defined by hard wall boundary conditions at both ends. A portion of this system's Hamiltonian spectrum was then calculated for the minimum energy configuration determined by $\phi_\mathrm{min}(B_y)$. The eigenvalues obtained from this Hamiltonian were then broadened using a Lorentzian profile of width $\epsilon = \Delta_0 / 100$ and plotted in Fig.~\ref{Fig2}(c). All eigenvalues are plotted, irrespective of the associated wavefunctions and their spatial localisation.

In all calculations, the model parameters used were as follows: $A_1=0.2$, $A_2=0.3$, $B_1=-0.2$, $B_2=-0.5$, $D_1=-0.05$, $D_2=0.15$, $M=-0.25$, $\mu_\mathrm{TI}=0.01$, $t=1.0$, $\mu_\mathrm{SC}=0.5$, $\Delta_0=0.025$, $V_\text{GT}=0.025$. The width of the nanowire was $W_\mathrm{TI}=18$, the widths of each of the superconducting electrodes was $W_\mathrm{SC}=6$, and the height of the device was $H=18$. These dimensions of the device cross-section were the same in the translationally-invariant and finite-length nanowire configurations. In the latter configuration, the length of the device was $L=150$. The length units in all cases are nanometers and the energy units are electronvolts.

{\bf  Acknowledgments:} 
This project has received funding from the European Research Council (ERC) under the European Union’s Horizon 2020 research and innovation program (Grant Agreement No. 741121) and was also funded by the Deutsche Forschungsgemeinschaft (DFG, German Research Foundation) under Germany’s Excellence Strategy - Cluster of Excellence Matter and Light for Quantum Computing (ML4Q) EXC 2004/2 - 390534769, as well as by the DFG under CRC 1238 - 277146847 (Subprojects A04 and B01). The authors also acknowledge support from the DFG Major Instrumentation Programme under project No. 544410649. EN acknowledges support by the Studienstiftung des deutschen Volkes. The research was funded in part by The Robert A. Welch Foundation, Grant \#L-E-0001-19921203.

{\bf Author contributions:} {Y.A. and J.S. conceived the project. M.B. grew the TI crystals. E.N. and J.S.  fabricated the devices, performed the experiments, and analyzed the data with inputs from Y.A. M.P. performed the theoretical modelling with help from L.F. M.G., and H.F.L. E.N. and H.F.L. wrote the manuscript with help from J.S., Y.A., and input from all authors.
}

{\bf Competing Interests:} The authors declare no competing interests.

{\bf Correspondence:} Correspondence and requests for materials should be addressed to Henry F. Legg and Yoichi Ando.

{\bf Data availability:} The data that support the findings of this study are available on Zenodo The data that support the findings of this study are available at the online depository zenodo with the identifier \href{https://doi.org/10.5281/zenodo.21988906}{10.5281/zenodo.21988906} and Supplementary Information.

\bibliography{nanoSQUID}

\end{document}


{\bf{\large Supplementary Information for ``Flux-tunable global and local superconductivity in a topological insulator nano-SQUID''}
}


\section{Experimental details and additional data}

\subsection{Sample geometry}

After completion of the transport measurements, the device dimensions were extracted from scanning electron microscopy (SEM) images to determine both length $L$ and width $W$ of the nanowires. The nanowire thickness was obtained using atomic force microscopy (AFM). A summary of the geometric parameters of devices A, B, and C is provided in Table \ref{table:Tab1}. Note that the nanowire width $W$ corresponds to the junction length.

\begin{table}[h]
\small
    \centering
    \begin{tabular}{ cl l l l}
    Device & $d$ (nm) & $W$ (nm) & $L$ ($\mu$m) \\ 
    \hline
    A & 21 & 70 & 1.9 \\  
    B & 36 & 70  & 1.3 \\  
    C & 16 & 80 & 2.5 \\
\end{tabular}
\caption{Nanowire dimensions for devices A--C.}
\label{table:Tab1}
\end{table}

\subsection{Magnetic-field alignment}
Reliable measurements in the presence of an axial magnetic field, $B_x$, parallel to the nanowire axis require careful alignment of the field direction. Most critical is an unintended out-of-plane component, $B_z$, which causes spatial variations of the superconducting phase across the junction. 

To calibrate the magnetic field alignment, Fraunhofer-type measurements were performed at finite nominal $B_x$. For each value, the corresponding compensation field $B_z$ was found by identifying the amount of field needed for the critical current to reach its maximum value. The required compensation directly reflects the spurious out-of-plane field component and was used to determine the correct field orientation. Further details of this procedure are given in Ref. \cite{Nikodem2025}.

In contrast, small deviations within the device plane were found to have a negligible effect on the measured transport data.

\subsection{Effects of measurement circuit}
The nonlocal conductance data shown in this work were corrected for the effects of measurement circuit arising from finite line impedance using the formalism developed in Ref. \cite{Martinez2021}. In particular, voltage-divider artifacts and systematic deviations introduced by the cryostat wiring and filtering stages were removed by transforming the measured conductance matrix into the conductance matrix at the device. The line resistances required were determined from two-terminal resistance measurements of the cryostat lines.

\subsection{Stability of the tunnel spectra}

\begin{figure}[b]
\centering
\includegraphics[width=0.8\textwidth]{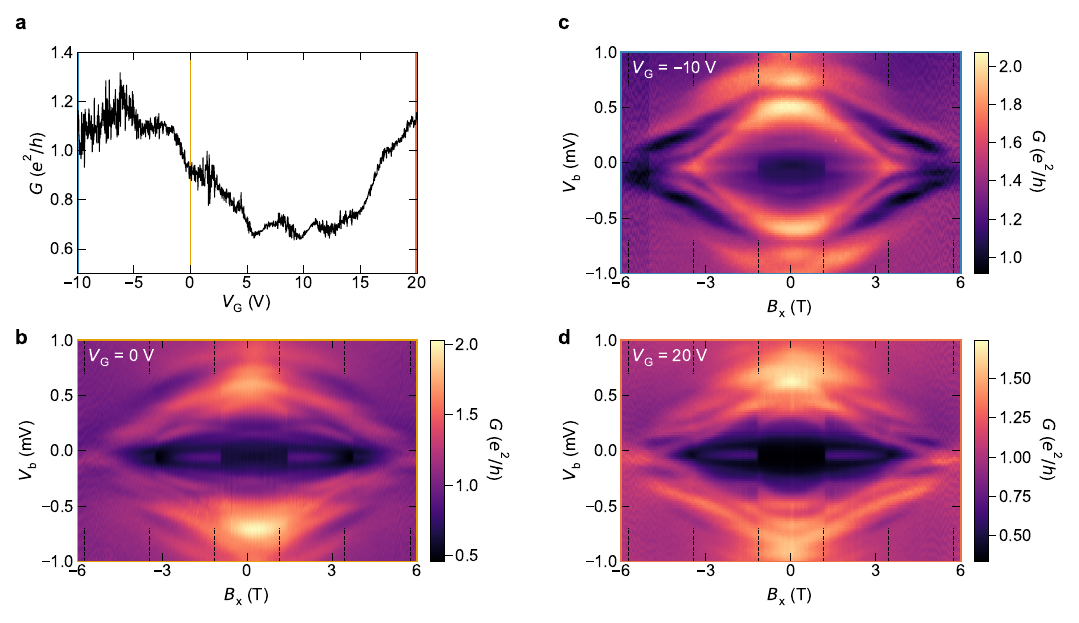}
\caption{\textbf{Gate-voltage dependence of the tunnel spectra in device A.} \textbf{a,} Differential conductance measured at a high bias voltage of 2 mV as a function of the globally applied back-gate voltage, $V_\text{G}$. The vertical dashed lines indicate the gate voltages at which the spectra shown in panels b, c, and d were acquired. \textbf{b, c, d,} Magnetic-field evolution of the spectrum at gate voltages $V_\text{G}=-10,0,20$ V.}
\label{FigS1}
\end{figure}

In Fig. 1g of the main text, we show the tunnel spectrum of device A in response to an applied bias voltage, $V_\text{b}$, at a gate voltage $V_\text{G} = 10$ V. To demonstrate that the sudden appearance of the zero-bias conductance peak in the expected topological regime is robust over a range of chemical potentials, we here show data taken at various gate voltages. Figure \ref{FigS1}a shows the conductance $G$ in zero magnetic field at a high bias voltage of $V_\text{b}=2$ mV, which essentially reflects the normal-state density of states, as a function of back-gate voltage $V_\text{G}$. The V-like shape of the LDOS with a minimum near $V_\text{G} = 10$ V may indicate that the Fermi level crosses the Dirac point around this gate voltage range.

At the gate voltages highlighted by dashed lines in Fig. \ref{FigS1}a, $V_\text{G} = -10$, 0, and 20~V, tunnel spectra were recorded as a function of in-plane magnetic field $B_x$. It becomes clear that the main feature of the spectra, namely the abrupt appearance of the zero-bias conductance anomaly at half-flux quantum, is very stable over the examined gate-voltage range, while higher-lying Andreev bound states are affected by the change in chemical potential.
Note that, in this device A, the gating range was limited due to a leaky gate dielectric and, thus, the risk of electrostatic discharge.

\subsection{Reproducibility in an additional device}

\begin{figure}[b]
\centering
\includegraphics[width=0.9\textwidth]{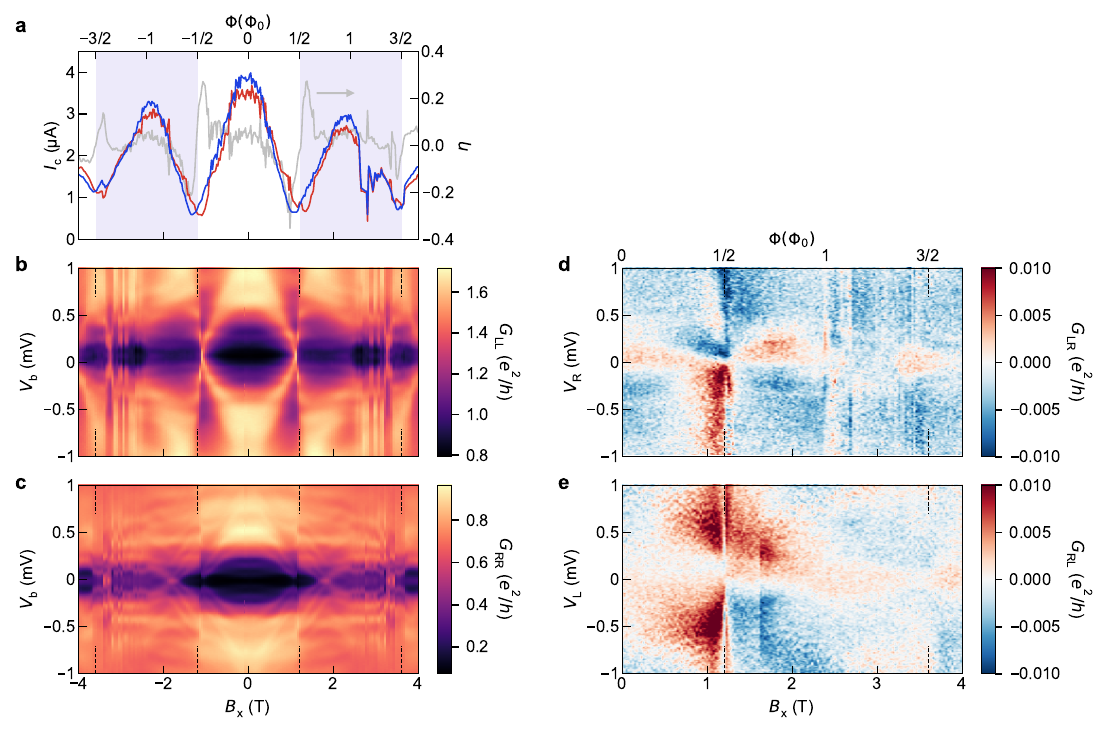}
\caption{\textbf{Simultaneous global and dual-local spectroscopy in device C.} \textbf{a,} Global critical current, $I_\text{c}$, and diode efficiency, $\eta$. \textbf{b, c,} Differential conductance spectra measured simultaneously at TJ1 and TJ2. \textbf{d, e,} Nonlocal conductance measurements demonstrating that some spectral features coincide with the boundaries of the expected topological phases.}
\label{FigS2}
\end{figure}

We were able to fabricate an additional device C with two functional tunnel probes similar to the devices presented in the main text. All key features are reproduced in this device. In Fig. \ref{FigS2}a, we show the critical current as a function of in-plane magnetic field $B_x$, along with the diode efficiency $\eta$, calculated from $I_\text{c}^+$ and $I_\text{c}^-$. Around a magnetic flux of $\Phi=h/4e$, the critical currents are minimal, and the diode polarity changes. At this flux value, discontinuities are observed in the magnetic-field-dependent spectra measured simultaneously at both tunnel junctions TJ1 and TJ2, although a zero-bias conductance peak was mostly absent in this device, as shown in Figs. \ref{FigS2}b and c. 

As in device B discussed in the main text, the presence of two tunnel probes enables nonlocal conductance measurements across the length of the nanowire, providing an additional probe of the global superconducting state. The nonlocal conductance exhibits abrupt spectral changes at magnetic fields corresponding to the expected first phase boundaries (Fig. \ref{FigS2}d, e). However, not all features can be associated with critical-current anomalies. In particular, additional spectral jumps occur at large magnetic fields. Due to the small cross-section of the nanowire compared to that of device B, the magnetic field corresponding to the second expected topological phase transition is already in a regime where vortex-related phenomena are expected to become prevalent. 

\begin{figure}[b]
\centering
\includegraphics[width=\textwidth]{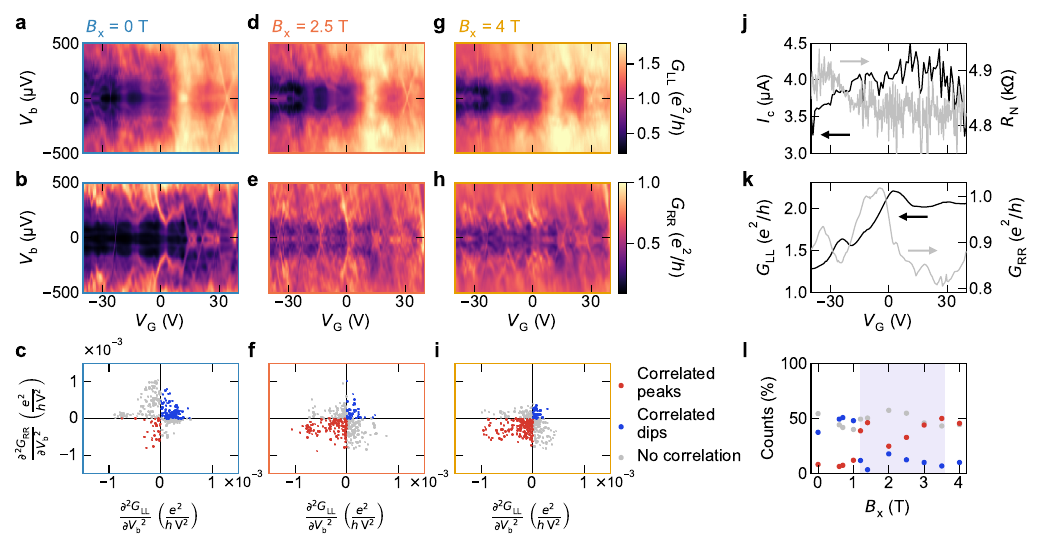}
\caption{\textbf{Gate-voltage dependence and end-to-end correlations in device C.} \textbf{a, b, d, e, g, h,} Differential conductance at TJ1 and TJ2 as a function of gate voltage, $V_\text{G}$, for trivial and topological flux regimes. \textbf{c, f, i,} End-to-end correlation analysis using the product of the the third derivative at zero bias. \textbf{j,} Gate-voltage dependence of the critical current (black), and the normal state resistance (gray). \textbf{k,} Gate-voltage dependence of the differential conductance of TJ1 (black) and TJ2 (gray). \textbf{l,} Relative fraction of gate voltages exhibiting correlated zero-bias conductance peaks (red), correlated zero-bias conductance dips (blue), or no correlation (grey) as a function of the magnetic field, $B_x$.}
\label{FigS6}
\end{figure}

To further investigate the emergence of correlated end-to-end states, we performed the same gate-dependent analysis as presented for device B in the main text. Figure \ref{FigS6}a and b show $G_{LL}$ and $G_{RR}$ measured at TJ1 and TJ2, respectively, as a function of $V_\text{G}$ and $V_\text{b}$ at zero magnetic field. Figure \ref{FigS6}d and e display the corresponding data at $B_x=2.5$ T, which is within the expected first topological regime, while Fig. \ref{FigS6}g and h show the spectra at $B_x=4$ T, which corresponds to the second expected trivial regime. To quantify the degree of end-to-end correlation, we performed the analysis as described in the main text. The results are presented in Fig. \ref{FigS6}c, f, i, and l. Similar to device B, we observe a strong increase of correlated zero-bias conductance peaks upon entering the first expected topological regime, but the number does not drop noticeably when reentering the expected trivial regime. 
Figure \ref{FigS6}j displays the critical current and normal-state resistance as a function of $V_\text{G}$, while Fig. \ref{FigS6}k shows the $V_\text{G}$ dependence of $G_{LL}$ and $G_{RR}$.

\subsection{Robustness of the diode effect}

\begin{figure}[b]
\centering
\includegraphics[width=0.9\textwidth]{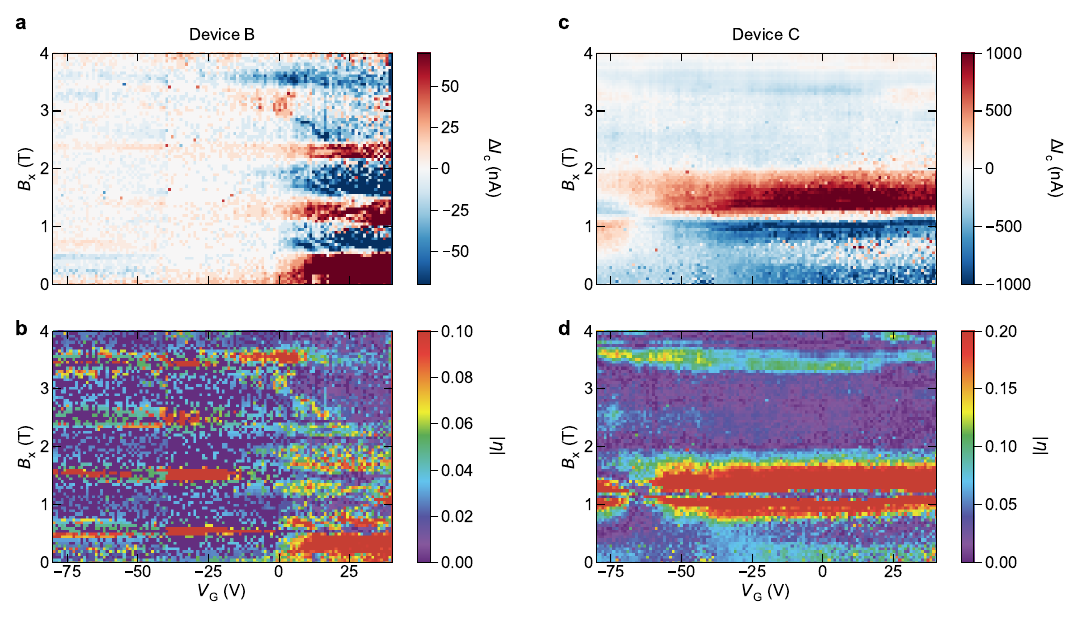}
\caption{\textbf{Tunability of the diode effect in devices B and C.} \textbf{a, b, c, d} Color mapping of $\Delta I_\mathrm{c} = I_\mathrm{c}^+-I_\mathrm{c}^-$ and the absolute value of the diode efficiency, $|\eta|$, as a function of $V_\mathrm{G}$ and $B_x$ in devices B and C.}
\label{FigS3}
\end{figure}

Figure \ref{FigS3} shows the superconducting diode effect in devices B and C as a function of $V_\text{G}$ and $B_x$. Panels a and c display the difference between the critical currents measured for positive and negative bias directions, $I_\text{c}^+-I_\text{c}^-$, plotted as a colour map in the $V_\text{G}$ vs $B_x$ plane. Panels b and d show the corresponding colour map of the absolute value of the diode efficiency. 

In device B, the sign reversals of the diode effect exhibit a largely regular periodicity as a function of $B_x$. In contrast, the behavior in device C becomes increasingly irregular at higher $B_x$, with the zero-crossing becoming less well-defined. This observation is consistent with the spectroscopic measurements, which also show a loss of regularity and the emergence of additional features in the high-field regime, suggesting a complex interplay between the mechanisms underlying the diode effect and additional vortex-related phenomena that become relevant in this regime.


\subsection{Gate-voltage dependence of the nonlocal conductance}

\begin{figure}[h]
\centering
\includegraphics[width=0.9\textwidth]{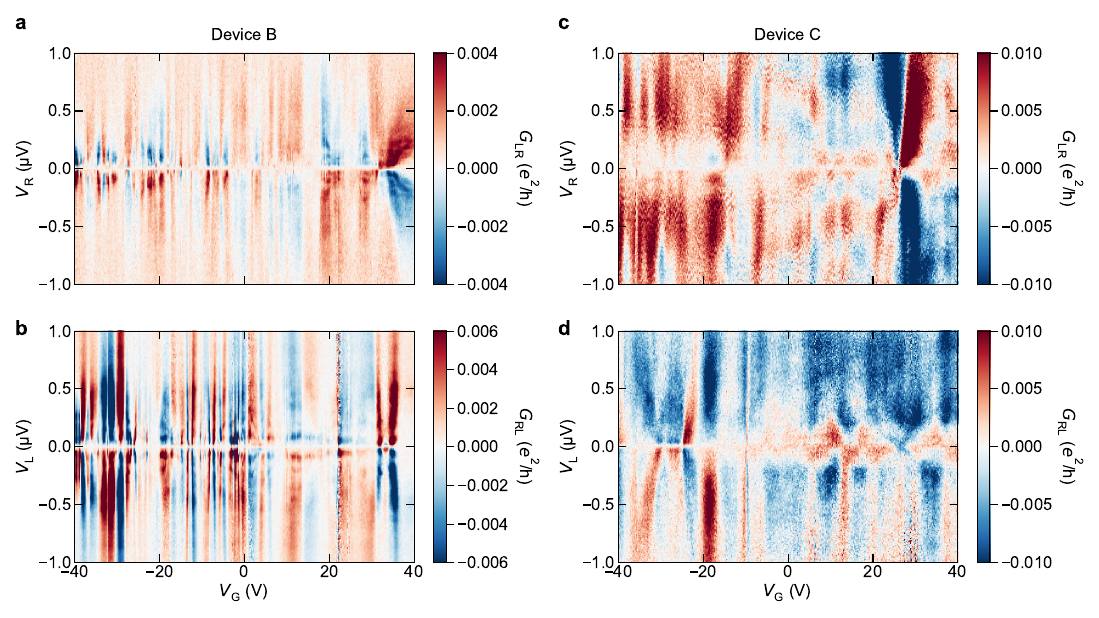}
\caption{\textbf{Gate-voltage dependence of the nonlocal conductance in devices B and C.} }
\label{FigS5}
\end{figure}

Figure \ref{FigS5} presents the nonlocal conductance $G_{LR}$ ($G_{RL}$) measured in devices B and C as a function of $V_\text{G}$ and bias voltages $V_\text{R}$ ($V_\text{L}$). In both devices, the low-bias nonlocal conductance exhibits frequent sign changes as the gate voltage is varied. The alternating positive and negative nonlocal response indicates a strong chemical-potential dependence of the nonlocal transport processes.

\bibliography{nanoSQUID}